\documentclass[proof]{WileyASNA-v1}

\usepackage[english]{babel}

\usepackage{microtype}
\usepackage{amsmath}
\usepackage{xspace}

\newcommand{\inte}{\textsl{INTEGRAL}\xspace}

\newcommand{\xmm}{\textsl{XMM-Newton}\xspace}
\newcommand{\nustar}{\textsl{NuSTAR}\xspace}
\newcommand{\xte}{\textsl{RXTE}\xspace}

\newcommand{\swift}{\textsl{Swift}\xspace}

\newcommand{\nh}{\ensuremath{{N}_\mathrm{H}}\xspace}

\newcommand{\snr}{S/R\xspace}

\newcommand{\gx}{GX~301$-$2\xspace}

\newcommand{\vela}{Vela~X-1\xspace}
\newcommand{\her}{Her~X-1\xspace}

\articletype{XMM-Newton Workshop 2018 Proceedings}%

\received{26 April 2016}
\revised{6 June 2016}
\accepted{6 June 2016}

\begin{document}

\title{Variability in High Mass X-ray Binaries}

\author[1]{Felix F\"urst}
\author[1]{Peter Kretschmar}

\author[2,3]{Victoria Grinberg}

\author[4,5]{Katja Pottschmidt}

\author[6]{J\"orn Wilms}


\author[6]{Matthias K\"uhnel}

\author[7]{Ileyk El Mellah}

\author[8]{Silvia Mart\'inez-N\'u\~nez}


\address[1]{European Space Astronomy Centre (ESAC), Science Operations Departement, 28692 Villanueva de la Ca\~nada, Madrid, Spain}

\address[2]{ESA European Space Research and Technology Centre (ESTEC), Keplerlaan 1, 2201 AZ Noordwijk, The Netherlands}

\address[3]{Institut für Astronomie und Astrophysik, Universit\"at T\"ubingen, Sand 1, 72076 T\"ubingen, Germany}

\address[4]{CRESST, Department of Physics, and Center for Space Science and Technology, UMBC, Baltimore, MD 21250, USA}
\address[5]{NASA Goddard Space Flight Center, Greenbelt, MD 20771, USA}

\address[6]{Dr. Karl-Remeis Sternwarte Bamberg \& ECAP, Sternwartstr. 7, 96049 Bamberg, Germany}

\address[7]{Centre for mathematical Plasma Astrophysics, Department of Mathematics, KU Leuven, Celestijnenlaan 200B, B-3001 Leuven, Belgium.}

\address[8]{Instituto de F\'isica de Cantabria (CSIC-Universidad de Cantabria), E-39005, Santander, Spain}

\corres{Felix F\"urst, \email{felix.fuerst@sciops.esa.int}}


\abstract{Strongly magnetised, accreting neutron stars show periodic and aperiodic variability on a large range of time-scales. By obtaining spectral and timing information on these different time-scales, we can obtain a closer look into the physics of accretion close to the neutron star and the properties of the accreted material.
One of  the most prominent time-scales are the strong pulsations, i.e., the rotation period of the neutron star itself. Over one rotation
our view of the accretion column and the X-ray producing region changes significantly. This allows us to sample different physical conditions within the column but at the same time requires that we have
viewing-angle resolved models to properly describe them.
In wind-fed high-mass X-ray binaries the main source of aperiodic variability is the clumpy stellar wind,
which leads to changes in accretion rate (i.e., luminosity) as well as absorption column. This
variability allows us to study the behavior of the accretion column as function of luminosity, as
well as to investigate the structure and physical properties of the wind, which we can
compare to winds in isolated stars.
}



\maketitle


\section{Introduction}
\label{sec:intro}

X-ray binaries, in particular accreting, magnetised neutron stars were one of the first X-ray sources discovered and identified through their strong pulsations \citep{giacconi71a}. Studies of their variability have therefore always been an integral part of their analysis. Studying the origin and effect of this variability will lead us to a more thorough understanding of their physics.
 In this paper we give a brief overview over the different types of variability we observe and how they can be exploited to learn more about the accreting system. We make no claim as to being a complete review but rather aim to show  the plethora of phenomena and quickly touch on some of their physical explanations. 

While this paper is focused on the timing properties, a brief introduction to the typical spectral properties of these systems is necessary to understand the implications of the variability. The bulk of the X-ray spectrum is produced in the accretion column close to the neutron star surface, where the material is channeled by the very strong magnetic field. The hot in-falling plasma can Compton up-scatter seed photons
 and produce a characteristic hard power-law ($\Gamma \ll 2$) spectrum with an exponential cutoff at high energies (with typical cutoff energies between 15--50\,keV, see, e.g., \citealt{white83a}). Typically this continuum is fit with phenomenological models, but recently advances on a physical modelling have been made \citep{becker05a, becker05b, becker07a, postnov15a,wolff16a}. At low energies the continuum is often attenuated by absorption, originating from the strong stellar wind or the inter-stellar medium. Additionally we might observe a thermal blackbody body component below $\sim$5\,keV, which might originate from the column itself or from an accretion disk.

An important element of the hard X-ray spectrum are  cyclotron resonant scattering features (CRSFs), which allow us to measure the magnetic field strength close to the surface of the  neutron star directly. They are formed by resonant scattering of photons with electrons quantised on Landau levels perpendicular to the magnetic field lines. As the energy of the Landau levels is directly proportional to the magnetic field strength, the energy of the observed CRSF gives us the magnetic field strength at the line forming region, following the ``12-$B$-12'' rule \citep[see, e.g.,][and references therein]{schwarm17a,schwarm17b}.

In the following Sections we discuss how these continuum parameters  vary as function of pulse phase, time, and/or flux, allowing us to measure different physical conditions in the accretion column close to the neutron star. We also discuss the  absorption variability, which occurs typically aperiodically or on orbital timescales. 

\section{Pulsations}
\label{sec:puls}

\subsection{Phenomenology}

Pulsations are caused by the rotation of the neutron star, with the  X-ray radiation from the accretion column moving through our line of sight.   Typically, the pulse period ranges from $\sim$1\,s \citep[e.g., Her~X-1,][]{tananbaum72a}  to thousands of seconds \citep[e.g., 4U~0114+65;][]{finley92a}. 

\begin{figure}
\begin{center}
\includegraphics[width=0.85\columnwidth]{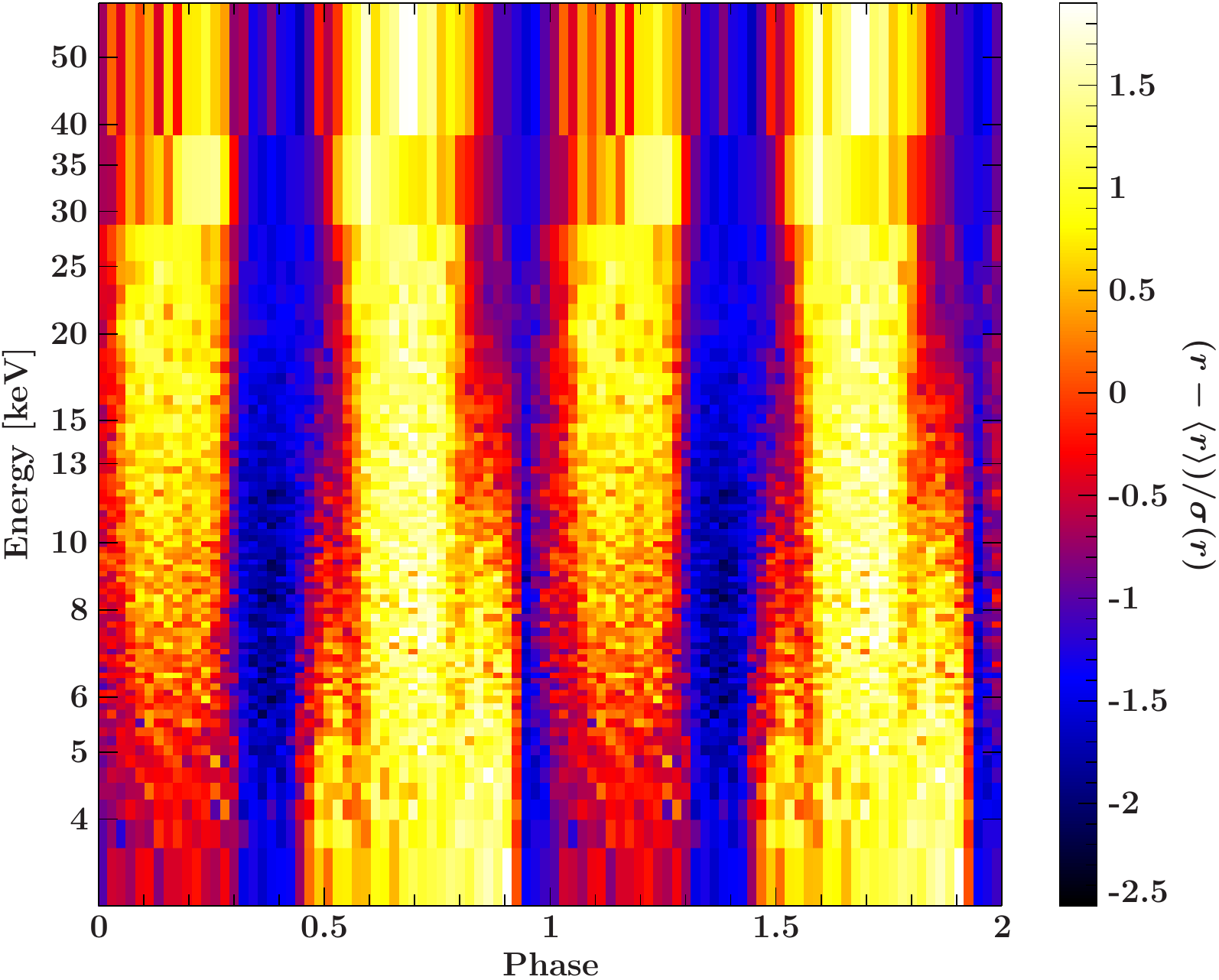}
\caption{Color-coded map of  pulse profiles of \gx as function of energy as observed with \nustar. The profiles were normalized according to their variance and are shown twice for clarity. Reproduced from \citet{gx301nustar}.}
\label{fig:ppergmap}
\end{center}
\end{figure}

Folding the light-curve on the observed pulse period leads to a pulse profile,  which shows the time-averaged flux as function of phase. These profiles are typically rather complex, with multiple sub-peaks and wide and narrow dips, and are strongly energy dependent (Fig.~\ref{fig:ppergmap} shows a typical example). The strong energy dependence already indicates that the observed X-ray spectrum changes significantly as function of pulse phase. 

Most sources show significant pulse-to-pulse variability, i.e., each individual pulse might deviate strongly from the average pulse profile. Sometimes this leads to missing single pulses \citep{goegues11a} or the vanishing of detectable pulsations altogether \citep{brumback18a}.
To circumvent the problem of strong pulse-to-pulse variability, different authors have used techniques like ``pulse-height resolved spectroscopy'' \citep{klochkov11a} and ``pulse-to-pulse" analysis \citep{gx301xmm}, however, at the expense of signal-to-noise ratio (\snr) and phase resolution.

\begin{figure}
\begin{center}
\includegraphics[width=0.95\columnwidth, trim={3.5cm 7.5cm 3.5cm 7cm}, clip]{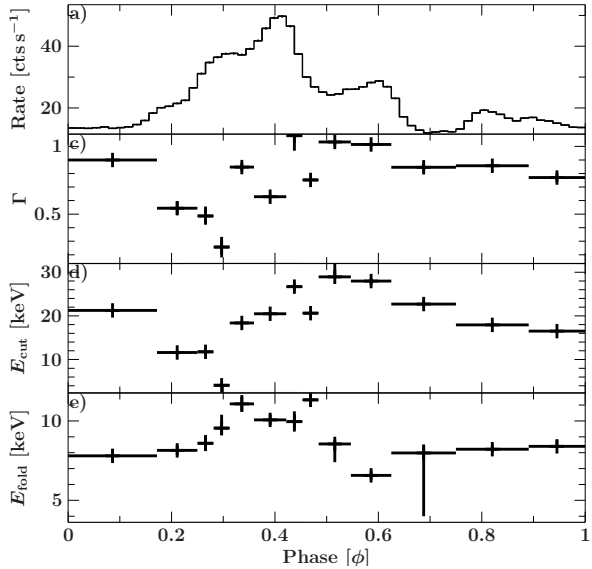}
\caption{Pulse profile (top panel) and phase dependence of the spectral parameters in \her. Reproduced from \citet{herx1}.}
\label{fig:pp_herx1}
\end{center}
\end{figure}

In most sources we see strong variation of the continuum parameters with pulse phase (photon-index, cutoff energy) and, where detected, also a strong variation of the CRSF parameters (in particular of the energy), as shown in Fig.~\ref{fig:pp_herx1}. In fact, in some sources the CRSF is only detected at a certain phase \citep[e.g., in KS~1947+300,][]{ks1947}. Mostly the variation is slow and smooth, however, there are also examples where the photon-index suddenly changes dramatically, over as short as 1\% of the phase \citep[e.g., EXO 2030+375,][]{ferrigno16a, exo2030}.

\subsection{Physical implications}

The variation as function of phase can be understood as viewing different parts of the accretion column (or different accretion columns) over the rotation of the neutron star. 
X-ray photons might leave the columns either along the magnetic field lines (``pencil beam'') or perpendicular to it (``fan beam''), or any combination thereof \citep{basko76a, meszaros85a}.
The physical conditions within the accretion column are not homogenous, but change with, e.g., height above the neutron star surface, and therefore we see spectral changes if our line-of-sight crosses different parts of the column. While this is qualitatively understood, it is very difficult to obtain quantitative measurements of the emission geometry or magnetic field configurations from these observations, due to the strong gravitational light-bending around the neutron star \citep[e.g.,][Falkner et al., subm]{sotani18a}. The observable flux from any part in the column changes dramatically for small changes in height and emission pattern, as the light-rays need to follow the Schwarzschild metric. Sadly, the emission pattern of accretion columns is not known (but see recent work by \citealt{postnov15a}) so that a reconstruction of the column geometry is very difficult. To make matters worse, reprocessing of the emitted X-rays on the neutron star surface can likely not be neglected \citep{mushtukov15a}, in particular because the in-falling plasma has velocities up to 0.7\,$c$ towards the neutron star, resulting in strong relativistic downwards beaming.

A recently developed theoretical model by Falkner et al. (2018, subm.) attempts to take all these effects into account and has been applied successfully to a few sources and their pulse phase-resolved behavior (Iwakiri et al., subm.).  In particular, using the energy variation of the CRSF in \gx, the model was used to measure the line-forming altitude, based on assumptions about the emission pattern and neutron star parameters \citep{gx301nustar}. However, the proposed solutions are not guaranteed to be the best or even unique.

Theoretically, by studying the pulse phase-resolved behavior, we will be able to constrain the emission pattern and magnetic field geometry, and together with physical models of the accretion column, obtain information about the compactness of the neutron star. If it is possible to constrain either mass or radius through independent measures (e.g., if the mass can be obtained from the mass function of the binary), these investigations will lead to new constraints on the equation of state of neutron stars.

\section{Orbital effects}
\label{sec:orbit}

In high-mass X-ray binary systems, orbital periods are found typically on the order of tens of days. The easiest way to determine the orbital period is through X-ray eclipses, which occur in high inclination systems \citep{falanga15a}.  

Be-systems, however, where the companion is  an O or B star with a circumstellar disk, mostly have eccentric orbits on the order of hundreds of days and rarely show eclipes \citep{corbet84a, knigge11a}. Many of these systems have eccentric orbits, resulting in  regular outbursts at periastron passage, which also can give a rough estimate of the orbital ephemeris \citep{priedhorsky83a}. However, the exact phases  and durations of outburst depend strongly on the activity of the donor star and outbursts do not necessarily happen close to periastron. For example in \gx and 4U~1907+09 so-called pre-periastron outbursts are observed, which occur around phase 0.8--0.9 \citep{leahy08a, kostka10a}.

\begin{figure}
\begin{center}
\includegraphics[width=0.99\columnwidth]{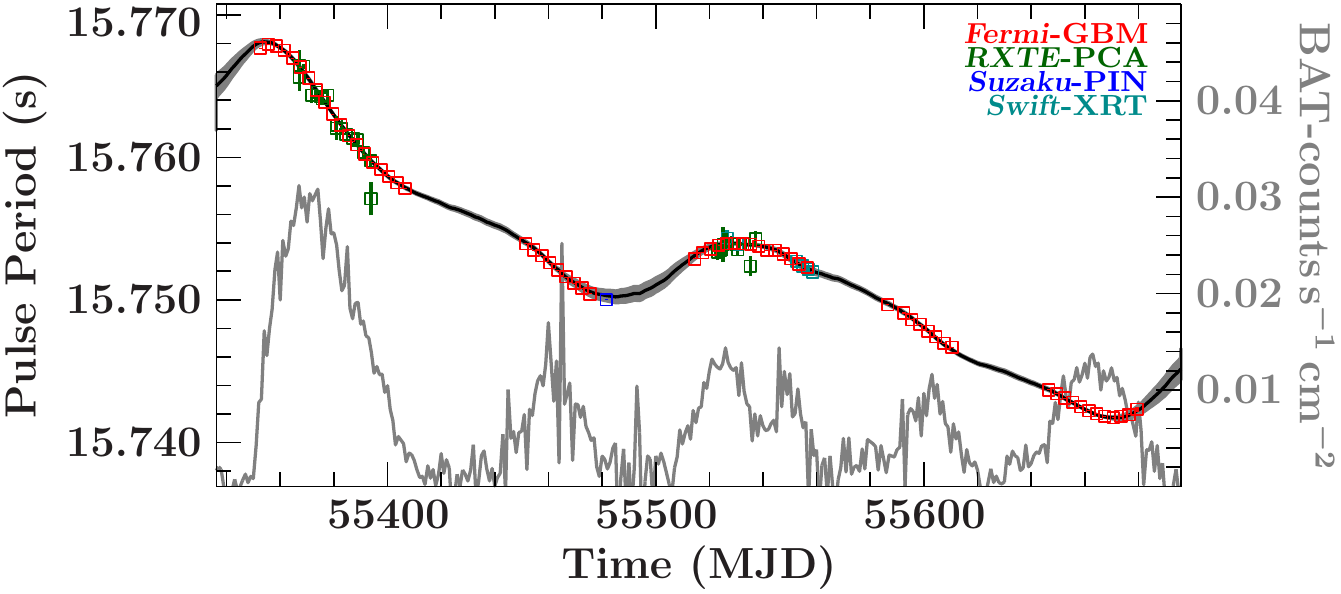}
\caption{Pulse period history of XTE~J1946+274, showing significant spin-up and variation as function of orbital phase. The gray line shows the \swift/BAT monitoring lichtcurve and the fit is based on  the model proposed by  \citet{ghosh79a}. Reproduced from \citet{marcu-cheatham15a}.}
\label{fig:j1946}
\end{center}
\end{figure}

The most accurate way to determine the orbital ephemeris is through pulse timing, i.e., by using the Doppler shift of the pulse period as function of orbit. This requires a precise measurement of the pulse period, which is possible for many galactic sources (see, e.g., the Fermi/GBM Pulsar Project\footnote{\url{https://gammaray.nsstc.nasa.gov/gbm/science/pulsars.html}}, \citealt{finger09a}). However, as these sources also change their pulse period due to varying accretion torque transferred onto the neutron star by the accreted matter, a good understanding of the coupling between the matter and the magnetic field is necessary. The seminal work by \citet{ghosh79a} is the basis for most torque models, and still widely applied today \citep[see Fig.~\ref{fig:j1946} for XTE~J1946+274,][] {marcu-cheatham15a}.

Besides a change in mass-accretion rate, the most readily observable effect of the orbital motion onto the X-ray spectrum is a changing absorption column. In super-giant systems, the compact object is embedded deeply in the stellar wind, and hence the absorption column increases as the neutron star gets closer to superior conjunction. 
In particular in eclipsing systems, like Vela~X-1 and 4U\,1700$-$37, where we view the orbit almost edge on, the absorption column is significantly increased in the second half of the orbit  \citep[e.g.,][]{sato86a, haberl89a, haberl90a}. 

While  variability of the absorption column is typically not the same in every orbit (Mart\'inez-N\'u\~nez in prep.), when averaging over many orbits, a stable pattern emerges 
 from which we can draw conclusions about the wind structure \citep{malacaria16a}.
In Be-systems, an inclined and precessing circumstellar disk might strongly influence the observed absorption column, as seen in GX~304$-$1 \citep{kuehnel17a}. These events, however, not only depend on  orbital phase, but also on the precessing period of the disk (which might be connected to a super-orbital period).



\section{Super-orbital periods}

Super-orbital periods are regular variations observed on time-scales longer than the orbital period, typically of the order of 10--100s days. They are usually associated with a precessing accretion disk, as observed in  systems like Her X-1 and SMC X-1 \citep{boynton80a, hickox05a}. The precession probably originates from a misalignment between the neutron star spin axis and the orbital axis or from uneven radiation pressure onto the disk \citep{iping90a}. 
It is possible that over the super-orbital changes also the mass transfer, and hence the accretion torque, which would be observable as a change in the spin-period \citep{staubert06a,dage18a}.
Similar super-orbital periods are also observed in ultra-luminous X-ray pulsars (ULXPs), e.g., NGC\,5907 ULX1 \citep{walton17a} or P13 \citep{motch14a, p13orb}, however, in the later system the period seems to be over 1500\,d. Hence, ULXPs might be connected to standard HMXBs, but a detailed discussion is beyond the scope of this paper.

In wind-accreting sources, i.e., standard super-giant systems, on the other hand the existence of a large accretion disk is unlikely, given the low angular momentum of the accreted matter and the strong magnetic field of the neutron star. However, \citet{corbet13a} discovered super-orbital periods in a number of super-giant systems, upending this assumption. So far, no clear explanation for these super-orbital periods has emerged. The formation of a temporary accretion disk is unlikely due to the clear presence of these features in the long-time monitoring light-curves of \xte/ASM or \swift/BAT. \citet{koengisberger06a} discuss the possibility that tidally-induced oscillations in the outer layer of the supergiant might modulate the mass loss rate locally. On the other hand \citet{bozzo17a} propose that an accretion stream due to co-rotating wind features could be responsible and \citet{middleton18a} suggested Lense-Thirring precession of the inner accretion flow. 

\section{Aperiodic variability}

One of the  currently most active fields concerning HMXBs and also the one were most progress has been made in recent years is aperiodic variability. This observed variability occurs on time-scales from seconds to hours or days, and is not correlated with the periodic variations in the source. It results from two effects: variability of the accretion rate and variability of the absorption column. Note that these variations occur mainly in wind-fed HMXBs and less in Be-systems.

\subsection{Phenomenology}

\begin{figure}
\begin{center}
\includegraphics[width=0.85\columnwidth]{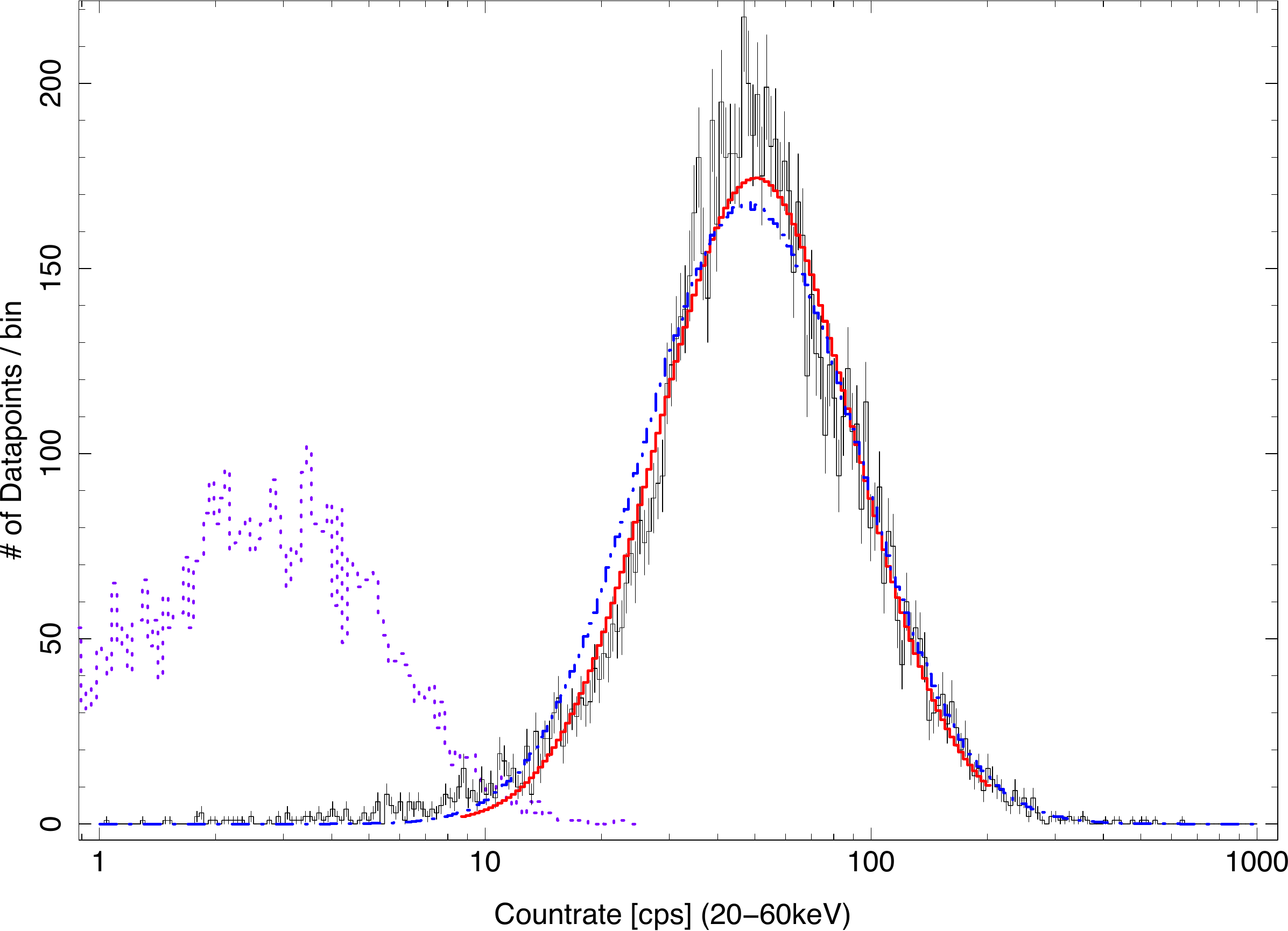}
\caption{Histogram of the 20--60\,keV count-rate of \vela as measured with \inte/ISGRI. The solid red line shows the best-fit log-normal distribution, while the dashed blue line shows the flux distribution from a simulated light-curve based on the power-spectrum. The purple dotted line shows the  background rate distribution. Reproduced from \citet{velastat}.}
\label{fig:velahisto}
\end{center}
\end{figure}

The variability of the accretion rate can be measured through the hard X-ray flux  of HMXBs, which is not influenced by variable absorption. The distribution of the flux shows a very clear log-normal distribution \citep[Figure~\ref{fig:velahisto},][]{velastat,  paizis14a}. 
Log-normal distributions are produced through multiplicative processes, e.g., we can imagine that large clumps in the wind are continuously separated into smaller clumps before being accreted onto the neutron star. This in turn allows us to estimate the clump masses that are being accreted, which are around $10^{19}$--$10^{21}$\,g \citep{velastat, walter07a, ducci09a}. However, these masses are significantly larger than the masses of clumps expected from wind models in single stars, indicating that more effects are in play in HMXBs \citep[for a detailed discussion on this topic see the review by][]{martinez17a}.

With modern instrumentation we now have the ability to study fluorescence and absorption lines created in the same material as the absorption, obtaining more information about its structure and ionisation state. For example, \citet{gx301xmm} have  used pulse-to-pulse spectroscopy to study \nh variability in \gx on time-scales as short as 685\,s (the pulse period) and found that the flux of the iron fluorescence lines mostly follows variation in absorption column. \citet{grinberg17a} used high-resolution grating spectroscopy to study \vela and could disentangle different states of absorption column density showing very different ionisation states. 

\subsection{Physical modeling}
\label{susec:physorb}
\citet{blondin90a} and \citet{blondin91a} developed a 2D hydrodynamical code to study the effects of the neutron star on a smooth wind, including the effects of the photo-ionization radiation from the neutron star on the wind acceleration. These authors found that indeed  semi-stable structures like an accretion wake are formed, which show enhanced density and absorption.
\citet{manousakis15a} extended this work, based on the same code, and found good agreement with the observed absorption variability over the whole orbit. As the mass of the neutron star is one of the most important factors in disrupting the smooth wind, \citet{manousakis12a} use the observed \nh variability to put constraints on the mass of the neutron star in IGR~J17252$-$3616. 

Simulations not only explain the absorption variability, but significant progress has also been made on describing the observed flux variability. \citet{manousakis15a} for example find that instabilities in the accretion flow, with low-density cavities forming, can explain the observed ``off-states'' in \vela \citep{kreykenbohm08a}. However, these results are still based on a smooth wind. \citet{oskinova12a} took a different approach, and based their simulations on a more sophisticated wind model for massive stars including clumps. They only model the wind in 1D and use a simple Bondi-Hoyle accretion radius \citep{bondi44a} and thus find that the variability produced by accretion of these clumps overestimates the observed flux variability significantly. 
The statistical representation of the clumpiness designed by \cite{ducci09a} led to a lower variability, and so does a fully multi-dimensional hydrodynamic treatment, which, however, is computationally expensive.


Recently, \citet{elmellah18a} presented 3D hydro-dynamic simulations, which use a realistic, clumpy wind which is disturbed by the presence of the neutron star. Applied to observational data of \vela, these simulations could not fully explain either the observed absorption variability \citep{grinberg17a} nor the observed flux variability, but showed that clearly further distortion of the wind by the neutron star occurs, potentially including the formation of a temporary accretion disk.
These results show that the X-ray ionizing feedback and the structure of the flow at the outer rim of the neutrons star magnetosphere play a major role and need to be determined (El Mellah et al., in prep).

\section{Conclusions}
\label{sec:conclusion}

In this work we have tried to summarize on a high level our current understanding of periodic and aperiodic variability in high-mass X-ray binaries. For more in-depth discussions of HMXBs we point the reader to the review by \citet{walter15a}, and to the review on stellar winds in these systems by \citet{martinez17a}.

We have discussed typical time-scales for pulsations, orbital periods, and super-orbital periods and how they allow us to view different physical processes in the system. With the discovery of ULXPs \citep{bachetti14a},  similar studies can now also be done in nearby galaxies  \citep[e.g.][]{ngc5907, p13orb}. This allows to extend the observable range of  luminosities and metallicities and see if accretion theories developed for local sources still hold.

We have made significant progress in recent years in our understanding of aperiodic variability, in particular with regards to the intrinsic X-ray flux and the variability of the absorption column. Through new observational capabilities and new hydro-dynamic simulations, we obtain more insight in the physical structure of the accreted medium, i.e., its density, mass, and ionization distribution. This distribution directly influences the observed flux variability. On the other hand, knowing the clump masses and sizes impacts implied stellar-mass loss rates, which in turn influence binary evolution scenarios.


\section*{Acknowledgments}
We thank the organisers of the \xmm 2018 science workshop  ``Time-Domain Astronomy: a High Energy View'' for the great conference and the opportunity to discuss our work.
Many results presented in these proceedings are based on work done by the X-MAG and X-WIND collaborations.



\begin{thebibliography}{}

\bibitem [\protect \citeauthoryear {%
{Bachetti}%
\ \protect \BOthers {.}}{%
{Bachetti}%
\ \protect \BOthers {.}}{%
{\protect \APACyear {2014}}%
}]{%
bachetti14a}
\APACinsertmetastar {%
bachetti14a}%
\begin{APACrefauthors}%
{Bachetti}, M.%
, {Harrison}, F\BPBI A.%
, {Walton}, D\BPBI J.%
\ et al.\end{APACrefauthors}%
\unskip\
\newblock
\APACrefYearMonthDay{2014}{{\APACmonth{07}}}{},
\newblock
\unskip
\newblock
\APACjournalVolNumPages{Nat}{514}{}{202}
\PrintBackRefs{\CurrentBib}

\bibitem [\protect \citeauthoryear {%
{Basko}%
\ \BBA {} {Sunyaev}%
}{%
{Basko}%
\ \BBA {} {Sunyaev}%
}{%
{\protect \APACyear {1976}}%
}]{%
basko76a}
\APACinsertmetastar {%
basko76a}%
\begin{APACrefauthors}%
{Basko}, M\BPBI M.%
\BCBT {}\ \BBA {} {Sunyaev}, R\BPBI A.%
\end{APACrefauthors}%
\unskip\
\newblock
\APACrefYearMonthDay{1976}{{\APACmonth{05}}}{},
\newblock
\unskip
\newblock
\APACjournalVolNumPages{MNRAS}{175}{}{395-417}
\PrintBackRefs{\CurrentBib}

\bibitem [\protect \citeauthoryear {%
{Becker}%
\ \BBA {} {Wolff}%
}{%
{Becker}%
\ \BBA {} {Wolff}%
}{%
{\protect \APACyear {2005}}%
{\protect \APACexlab {{\protect \BCnt {1}}}}}]{%
becker05a}
\APACinsertmetastar {%
becker05a}%
\begin{APACrefauthors}%
{Becker}, P\BPBI A.%
\BCBT {}\ \BBA {} {Wolff}, M\BPBI T.%
\end{APACrefauthors}%
\unskip\
\newblock
\APACrefYearMonthDay{2005{\protect \BCnt {1}}}{{\APACmonth{03}}}{},
\newblock
\unskip
\newblock
\APACjournalVolNumPages{ApJ}{621}{}{L45-L48}
\newblock
\PrintBackRefs{\CurrentBib}

\bibitem [\protect \citeauthoryear {%
{Becker}%
\ \BBA {} {Wolff}%
}{%
{Becker}%
\ \BBA {} {Wolff}%
}{%
{\protect \APACyear {2005}}%
{\protect \APACexlab {{\protect \BCnt {2}}}}}]{%
becker05b}
\APACinsertmetastar {%
becker05b}%
\begin{APACrefauthors}%
{Becker}, P\BPBI A.%
\BCBT {}\ \BBA {} {Wolff}, M\BPBI T.%
\end{APACrefauthors}%
\unskip\
\newblock
\APACrefYearMonthDay{2005{\protect \BCnt {2}}}{{\APACmonth{09}}}{},
\newblock
\unskip
\newblock
\APACjournalVolNumPages{ApJ}{630}{}{465-488}
\newblock
\PrintBackRefs{\CurrentBib}

\bibitem [\protect \citeauthoryear {%
{Becker}%
\ \BBA {} {Wolff}%
}{%
{Becker}%
\ \BBA {} {Wolff}%
}{%
{\protect \APACyear {2007}}%
}]{%
becker07a}
\APACinsertmetastar {%
becker07a}%
\begin{APACrefauthors}%
{Becker}, P\BPBI A.%
\BCBT {}\ \BBA {} {Wolff}, M\BPBI T.%
\end{APACrefauthors}%
\unskip\
\newblock
\APACrefYearMonthDay{2007}{{\APACmonth{01}}}{},
\newblock
\unskip
\newblock
\APACjournalVolNumPages{ApJ}{654}{}{435-457}
\newblock
\PrintBackRefs{\CurrentBib}

\bibitem [\protect \citeauthoryear {%
Blondin%
, Kallman%
, Fryxell%
\BCBL {}\ \BBA {} Taam%
}{%
Blondin%
\ \protect \BOthers {.}}{%
{\protect \APACyear {1990}}%
}]{%
blondin90a}
\APACinsertmetastar {%
blondin90a}%
\begin{APACrefauthors}%
Blondin, J\BPBI M.%
, Kallman, T\BPBI R.%
, Fryxell, B\BPBI A.%
\BCBL {}\ \BBA {} Taam, R\BPBI E.%
\end{APACrefauthors}%
\unskip\
\newblock
\APACrefYearMonthDay{1990}{{\APACmonth{06}}}{},
\newblock
\unskip
\newblock
\APACjournalVolNumPages{ApJ}{356}{}{591--608}
\PrintBackRefs{\CurrentBib}

\bibitem [\protect \citeauthoryear {%
Blondin%
, Stevens%
\BCBL {}\ \BBA {} Kallman%
}{%
Blondin%
\ \protect \BOthers {.}}{%
{\protect \APACyear {1991}}%
}]{%
blondin91a}
\APACinsertmetastar {%
blondin91a}%
\begin{APACrefauthors}%
Blondin, J\BPBI M.%
, Stevens, I\BPBI R.%
\BCBL {}\ \BBA {} Kallman, T\BPBI R.%
\end{APACrefauthors}%
\unskip\
\newblock
\APACrefYearMonthDay{1991}{{\APACmonth{04}}}{},
\newblock
\unskip
\newblock
\APACjournalVolNumPages{ApJ}{371}{}{684--695}
\PrintBackRefs{\CurrentBib}

\bibitem [\protect \citeauthoryear {%
Bondi%
\ \BBA {} Hoyle%
}{%
Bondi%
\ \BBA {} Hoyle%
}{%
{\protect \APACyear {1944}}%
}]{%
bondi44a}
\APACinsertmetastar {%
bondi44a}%
\begin{APACrefauthors}%
Bondi, H.%
\BCBT {}\ \BBA {} Hoyle, F.%
\end{APACrefauthors}%
\unskip\
\newblock
\APACrefYearMonthDay{1944}{}{},
\newblock
\unskip
\newblock
\APACjournalVolNumPages{MNRAS}{104}{}{273}
\PrintBackRefs{\CurrentBib}

\bibitem [\protect \citeauthoryear {%
{Boynton}%
, {Crosa}%
\BCBL {}\ \BBA {} {Deeter}%
}{%
{Boynton}%
\ \protect \BOthers {.}}{%
{\protect \APACyear {1980}}%
}]{%
boynton80a}
\APACinsertmetastar {%
boynton80a}%
\begin{APACrefauthors}%
{Boynton}, P\BPBI E.%
, {Crosa}, L\BPBI M.%
\BCBL {}\ \BBA {} {Deeter}, J\BPBI E.%
\end{APACrefauthors}%
\unskip\
\newblock
\APACrefYearMonthDay{1980}{{\APACmonth{04}}}{},
\newblock
\unskip
\newblock
\APACjournalVolNumPages{ApJ}{237}{}{169-174}
\PrintBackRefs{\CurrentBib}

\bibitem [\protect \citeauthoryear {%
{Bozzo}%
, {Oskinova}%
, {Lobel}%
\BCBL {}\ \BBA {} {Hamann}%
}{%
{Bozzo}%
\ \protect \BOthers {.}}{%
{\protect \APACyear {2017}}%
}]{%
bozzo17a}
\APACinsertmetastar {%
bozzo17a}%
\begin{APACrefauthors}%
{Bozzo}, E.%
, {Oskinova}, L.%
, {Lobel}, A.%
\BCBL {}\ \BBA {} {Hamann}, W\BHBI R.%
\end{APACrefauthors}%
\unskip\
\newblock
\APACrefYearMonthDay{2017}{{\APACmonth{10}}}{},
\newblock
\unskip
\newblock
\APACjournalVolNumPages{A\&A}{606}{}{L10}
\newblock
\PrintBackRefs{\CurrentBib}

\bibitem [\protect \citeauthoryear {%
{Brumback}%
\ \protect \BOthers {.}}{%
{Brumback}%
\ \protect \BOthers {.}}{%
{\protect \APACyear {2018}}%
}]{%
brumback18a}
\APACinsertmetastar {%
brumback18a}%
\begin{APACrefauthors}%
{Brumback}, M\BPBI C.%
, {Hickox}, R\BPBI C.%
, {Bachetti}, M.%
\ et al.\end{APACrefauthors}%
\unskip\
\newblock
\APACrefYearMonthDay{2018}{{\APACmonth{07}}}{},
\newblock
\unskip
\newblock
\APACjournalVolNumPages{ApJL}{861}{}{L7}
\newblock
\PrintBackRefs{\CurrentBib}

\bibitem [\protect \citeauthoryear {%
{Corbet}%
}{%
{Corbet}%
}{%
{\protect \APACyear {1984}}%
}]{%
corbet84a}
\APACinsertmetastar {%
corbet84a}%
\begin{APACrefauthors}%
{Corbet}, R\BPBI H\BPBI D.%
\end{APACrefauthors}%
\unskip\
\newblock
\APACrefYearMonthDay{1984}{{\APACmonth{12}}}{},
\newblock
\unskip
\newblock
\APACjournalVolNumPages{A\&A}{141}{}{91-93}
\PrintBackRefs{\CurrentBib}

\bibitem [\protect \citeauthoryear {%
{Corbet}%
\ \BBA {} {Krimm}%
}{%
{Corbet}%
\ \BBA {} {Krimm}%
}{%
{\protect \APACyear {2013}}%
}]{%
corbet13a}
\APACinsertmetastar {%
corbet13a}%
\begin{APACrefauthors}%
{Corbet}, R\BPBI H\BPBI D.%
\BCBT {}\ \BBA {} {Krimm}, H\BPBI A.%
\end{APACrefauthors}%
\unskip\
\newblock
\APACrefYearMonthDay{2013}{{\APACmonth{11}}}{},
\newblock
\unskip
\newblock
\APACjournalVolNumPages{ApJ}{778}{}{45}
\newblock
\PrintBackRefs{\CurrentBib}

\bibitem [\protect \citeauthoryear {%
{Dage}%
, {Clarkson}%
, {Charles}%
, {Laycock}%
\BCBL {}\ \BBA {} {Shih}%
}{%
{Dage}%
\ \protect \BOthers {.}}{%
{\protect \APACyear {2018}}%
}]{%
dage18a}
\APACinsertmetastar {%
dage18a}%
\begin{APACrefauthors}%
{Dage}, K\BPBI C.%
, {Clarkson}, W\BPBI I.%
, {Charles}, P\BPBI A.%
, {Laycock}, S\BPBI G\BPBI T.%
\BCBL {}\ \BBA {} {Shih}, I\BHBI C.%
\end{APACrefauthors}%
\unskip\
\newblock
\APACrefYearMonthDay{2018}{{\APACmonth{09}}}{},
\newblock
\unskip
\newblock
\APACjournalVolNumPages{MNRAS}{}{}{}
\newblock
\PrintBackRefs{\CurrentBib}

\bibitem [\protect \citeauthoryear {%
{Ducci}%
, {Sidoli}%
, {Mereghetti}%
, {Paizis}%
\BCBL {}\ \BBA {} {Romano}%
}{%
{Ducci}%
\ \protect \BOthers {.}}{%
{\protect \APACyear {2009}}%
}]{%
ducci09a}
\APACinsertmetastar {%
ducci09a}%
\begin{APACrefauthors}%
{Ducci}, L.%
, {Sidoli}, L.%
, {Mereghetti}, S.%
, {Paizis}, A.%
\BCBL {}\ \BBA {} {Romano}, P.%
\end{APACrefauthors}%
\unskip\
\newblock
\APACrefYearMonthDay{2009}{{\APACmonth{10}}}{},
\newblock
\unskip
\newblock
\APACjournalVolNumPages{MNRAS}{398}{}{2152-2165}
\newblock
\PrintBackRefs{\CurrentBib}

\bibitem [\protect \citeauthoryear {%
{El Mellah}%
, {Sundqvist}%
\BCBL {}\ \BBA {} {Keppens}%
}{%
{El Mellah}%
\ \protect \BOthers {.}}{%
{\protect \APACyear {2018}}%
}]{%
elmellah18a}
\APACinsertmetastar {%
elmellah18a}%
\begin{APACrefauthors}%
{El Mellah}, I.%
, {Sundqvist}, J\BPBI O.%
\BCBL {}\ \BBA {} {Keppens}, R.%
\end{APACrefauthors}%
\unskip\
\newblock
\APACrefYearMonthDay{2018}{{\APACmonth{04}}}{},
\newblock
\unskip
\newblock
\APACjournalVolNumPages{MNRAS}{475}{}{3240-3252}
\PrintBackRefs{\CurrentBib}

\bibitem [\protect \citeauthoryear {%
{Falanga}%
\ \protect \BOthers {.}}{%
{Falanga}%
\ \protect \BOthers {.}}{%
{\protect \APACyear {2015}}%
}]{%
falanga15a}
\APACinsertmetastar {%
falanga15a}%
\begin{APACrefauthors}%
{Falanga}, M.%
, {Bozzo}, E.%
, {Lutovinov}, A.%
, {Bonnet-Bidaud}, J\BPBI M.%
, {Fetisova}, Y.%
\BCBL {}\ \BBA {} {Puls}, J.%
\end{APACrefauthors}%
\unskip\
\newblock
\APACrefYearMonthDay{2015}{{\APACmonth{05}}}{},
\newblock
\unskip
\newblock
\APACjournalVolNumPages{A\&A}{577}{}{A130}
\newblock
\PrintBackRefs{\CurrentBib}

\bibitem [\protect \citeauthoryear {%
{Ferrigno}%
\ \protect \BOthers {.}}{%
{Ferrigno}%
\ \protect \BOthers {.}}{%
{\protect \APACyear {2016}}%
}]{%
ferrigno16a}
\APACinsertmetastar {%
ferrigno16a}%
\begin{APACrefauthors}%
{Ferrigno}, C.%
, {Pjanka}, P.%
, {Bozzo}, E.%
, {Klochkov}, D.%
, {Ducci}, L.%
\BCBL {}\ \BBA {} {Zdziarski}, A\BPBI A.%
\end{APACrefauthors}%
\unskip\
\newblock
\APACrefYearMonthDay{2016}{{\APACmonth{09}}}{},
\newblock
\unskip
\newblock
\APACjournalVolNumPages{A\&A}{593}{}{A105}
\newblock
\PrintBackRefs{\CurrentBib}

\bibitem [\protect \citeauthoryear {%
{Finger}%
\ \protect \BOthers {.}}{%
{Finger}%
\ \protect \BOthers {.}}{%
{\protect \APACyear {2009}}%
}]{%
finger09a}
\APACinsertmetastar {%
finger09a}%
\begin{APACrefauthors}%
{Finger}, M\BPBI H.%
, {Beklen}, E.%
, {Narayana Bhat}, P.%
\ et al.\end{APACrefauthors}%
\unskip\
\newblock
\APACrefYearMonthDay{2009}{{\APACmonth{12}}}{},
\newblock
{\BBOQ}\APACrefatitle {{Long-term Monitoring of Accreting Pulsars with Fermi
  GBM}} {{Long-term Monitoring of Accreting Pulsars with Fermi GBM}}.{\BBCQ}
\newblock
\BIn{} \APACrefbtitle {{Proc. of the 2009 Fermi Symposium, eConf Proceedings
  C091122}.} {{Proc. of the 2009 Fermi Symposium, eConf Proceedings C091122}.}
\newblock
\APACaddressPublisher{}{{arXiv:0912.3847}}.
\PrintBackRefs{\CurrentBib}

\bibitem [\protect \citeauthoryear {%
{Finley}%
, {Belloni}%
\BCBL {}\ \BBA {} {Cassinelli}%
}{%
{Finley}%
\ \protect \BOthers {.}}{%
{\protect \APACyear {1992}}%
}]{%
finley92a}
\APACinsertmetastar {%
finley92a}%
\begin{APACrefauthors}%
{Finley}, J\BPBI P.%
, {Belloni}, T.%
\BCBL {}\ \BBA {} {Cassinelli}, J\BPBI P.%
\end{APACrefauthors}%
\unskip\
\newblock
\APACrefYearMonthDay{1992}{{\APACmonth{09}}}{},
\newblock
\unskip
\newblock
\APACjournalVolNumPages{A\&A}{262}{}{L25-L28}
\PrintBackRefs{\CurrentBib}

\bibitem [\protect \citeauthoryear {%
{F{\"u}rst}%
\ \protect \BOthers {.}}{%
{F{\"u}rst}%
\ \protect \BOthers {.}}{%
{\protect \APACyear {2018}}%
}]{%
gx301nustar}
\APACinsertmetastar {%
gx301nustar}%
\begin{APACrefauthors}%
{F{\"u}rst}, F.%
, {Falkner}, S.%
, {Marcu-Cheatham}, D.%
\ et al.\end{APACrefauthors}%
\unskip\
\newblock
\APACrefYearMonthDay{2018}{{\APACmonth{09}}}{},
\newblock
\unskip
\newblock
\APACjournalVolNumPages{A\&A in press., arXiv:1809.05691}{}{}{}
\PrintBackRefs{\CurrentBib}

\bibitem [\protect \citeauthoryear {%
F{\"u}rst%
\ \protect \BOthers {.}}{%
F{\"u}rst%
\ \protect \BOthers {.}}{%
{\protect \APACyear {2013}}%
}]{%
herx1}
\APACinsertmetastar {%
herx1}%
\begin{APACrefauthors}%
F{\"u}rst, F.%
, {Grefenstette}, B\BPBI W.%
, {Staubert}, R.%
\ et al.\end{APACrefauthors}%
\unskip\
\newblock
\APACrefYearMonthDay{2013}{{\APACmonth{12}}}{},
\newblock
\unskip
\newblock
\APACjournalVolNumPages{ApJ}{778}{}{69}
\PrintBackRefs{\CurrentBib}

\bibitem [\protect \citeauthoryear {%
F{\"u}rst%
, {Kretschmar}%
\BCBL {}\ \protect \BOthers {.}}{%
F{\"u}rst%
, {Kretschmar}%
\BCBL {}\ \protect \BOthers {.}}{%
{\protect \APACyear {2017}}%
}]{%
exo2030}
\APACinsertmetastar {%
exo2030}%
\begin{APACrefauthors}%
F{\"u}rst, F.%
, {Kretschmar}, P.%
, {Kajava}, J\BPBI J\BPBI E.%
\ et al.\end{APACrefauthors}%
\unskip\
\newblock
\APACrefYearMonthDay{2017}{{\APACmonth{10}}}{},
\newblock
\unskip
\newblock
\APACjournalVolNumPages{A\&A}{606}{}{A89}
\newblock
\PrintBackRefs{\CurrentBib}

\bibitem [\protect \citeauthoryear {%
{F{\"u}rst}%
\ \protect \BOthers {.}}{%
{F{\"u}rst}%
\ \protect \BOthers {.}}{%
{\protect \APACyear {2010}}%
}]{%
velastat}
\APACinsertmetastar {%
velastat}%
\begin{APACrefauthors}%
{F{\"u}rst}, F.%
, {Kreykenbohm}, I.%
, {Pottschmidt}, K.%
\ et al.\end{APACrefauthors}%
\unskip\
\newblock
\APACrefYearMonthDay{2010}{{\APACmonth{09}}}{},
\newblock
\unskip
\newblock
\APACjournalVolNumPages{A\&A}{519}{}{A37}
\PrintBackRefs{\CurrentBib}

\bibitem [\protect \citeauthoryear {%
F{\"u}rst%
\ \protect \BOthers {.}}{%
F{\"u}rst%
\ \protect \BOthers {.}}{%
{\protect \APACyear {2014}}%
}]{%
ks1947}
\APACinsertmetastar {%
ks1947}%
\begin{APACrefauthors}%
F{\"u}rst, F.%
, {Pottschmidt}, K.%
, {Wilms}, J.%
\ et al.\end{APACrefauthors}%
\unskip\
\newblock
\APACrefYearMonthDay{2014}{{\APACmonth{04}}}{},
\newblock
\unskip
\newblock
\APACjournalVolNumPages{ApJL}{784}{}{L40}
\newblock
\PrintBackRefs{\CurrentBib}

\bibitem [\protect \citeauthoryear {%
F{\"u}rst%
\ \protect \BOthers {.}}{%
F{\"u}rst%
\ \protect \BOthers {.}}{%
{\protect \APACyear {2011}}%
}]{%
gx301xmm}
\APACinsertmetastar {%
gx301xmm}%
\begin{APACrefauthors}%
F{\"u}rst, F.%
, Suchy, S.%
, Kreykenbohm, I.%
\ et al.\end{APACrefauthors}%
\unskip\
\newblock
\APACrefYearMonthDay{2011}{}{},
\newblock
\unskip
\newblock
\APACjournalVolNumPages{A\&A}{535}{}{A9}
\PrintBackRefs{\CurrentBib}

\bibitem [\protect \citeauthoryear {%
F{\"u}rst%
\ \protect \BOthers {.}}{%
F{\"u}rst%
\ \protect \BOthers {.}}{%
{\protect \APACyear {2018}}%
}]{%
p13orb}
\APACinsertmetastar {%
p13orb}%
\begin{APACrefauthors}%
F{\"u}rst, F.%
, {Walton}, D\BPBI J.%
, {Heida}, M.%
\ et al.\end{APACrefauthors}%
\unskip\
\newblock
\APACrefYearMonthDay{2018}{{\APACmonth{09}}}{},
\newblock
\unskip
\newblock
\APACjournalVolNumPages{A\&A}{616}{}{A186}
\PrintBackRefs{\CurrentBib}

\bibitem [\protect \citeauthoryear {%
F{\"u}rst%
, {Walton}%
\BCBL {}\ \protect \BOthers {.}}{%
F{\"u}rst%
, {Walton}%
\BCBL {}\ \protect \BOthers {.}}{%
{\protect \APACyear {2017}}%
}]{%
ngc5907}
\APACinsertmetastar {%
ngc5907}%
\begin{APACrefauthors}%
F{\"u}rst, F.%
, {Walton}, D\BPBI J.%
, {Stern}, D.%
\ et al.\end{APACrefauthors}%
\unskip\
\newblock
\APACrefYearMonthDay{2017}{{\APACmonth{01}}}{},
\newblock
\unskip
\newblock
\APACjournalVolNumPages{ApJ}{834}{}{77}
\newblock
\PrintBackRefs{\CurrentBib}

\bibitem [\protect \citeauthoryear {%
{Ghosh}%
\ \BBA {} {Lamb}%
}{%
{Ghosh}%
\ \BBA {} {Lamb}%
}{%
{\protect \APACyear {1979}}%
}]{%
ghosh79a}
\APACinsertmetastar {%
ghosh79a}%
\begin{APACrefauthors}%
{Ghosh}, P.%
\BCBT {}\ \BBA {} {Lamb}, F\BPBI K.%
\end{APACrefauthors}%
\unskip\
\newblock
\APACrefYearMonthDay{1979}{{\APACmonth{08}}}{},
\newblock
\unskip
\newblock
\APACjournalVolNumPages{ApJ}{232}{}{259-276}
\newblock
\PrintBackRefs{\CurrentBib}

\bibitem [\protect \citeauthoryear {%
{Giacconi}%
, {Gursky}%
, {Kellogg}%
, {Schreier}%
\BCBL {}\ \BBA {} {Tananbaum}%
}{%
{Giacconi}%
\ \protect \BOthers {.}}{%
{\protect \APACyear {1971}}%
}]{%
giacconi71a}
\APACinsertmetastar {%
giacconi71a}%
\begin{APACrefauthors}%
{Giacconi}, R.%
, {Gursky}, H.%
, {Kellogg}, E.%
, {Schreier}, E.%
\BCBL {}\ \BBA {} {Tananbaum}, H.%
\end{APACrefauthors}%
\unskip\
\newblock
\APACrefYearMonthDay{1971}{{\APACmonth{07}}}{},
\newblock
\unskip
\newblock
\APACjournalVolNumPages{ApJ}{167}{}{L67}
\PrintBackRefs{\CurrentBib}

\bibitem [\protect \citeauthoryear {%
G\"{o}\u{g}\"{u}\c{s}%
, Kreykenbohm%
\BCBL {}\ \BBA {} Belloni%
}{%
G\"{o}\u{g}\"{u}\c{s}%
\ \protect \BOthers {.}}{%
{\protect \APACyear {2011}}%
}]{%
goegues11a}
\APACinsertmetastar {%
goegues11a}%
\begin{APACrefauthors}%
G\"{o}\u{g}\"{u}\c{s}, E.%
, Kreykenbohm, I.%
\BCBL {}\ \BBA {} Belloni, T\BPBI M.%
\end{APACrefauthors}%
\unskip\
\newblock
\APACrefYearMonthDay{2011}{{\APACmonth{01}}}{},
\newblock
\unskip
\newblock
\APACjournalVolNumPages{A\&A}{525}{}{L6}
\PrintBackRefs{\CurrentBib}

\bibitem [\protect \citeauthoryear {%
{Grinberg}%
\ \protect \BOthers {.}}{%
{Grinberg}%
\ \protect \BOthers {.}}{%
{\protect \APACyear {2017}}%
}]{%
grinberg17a}
\APACinsertmetastar {%
grinberg17a}%
\begin{APACrefauthors}%
{Grinberg}, V.%
, {Hell}, N.%
, {El Mellah}, I.%
\ et al.\end{APACrefauthors}%
\unskip\
\newblock
\APACrefYearMonthDay{2017}{{\APACmonth{12}}}{},
\newblock
\unskip
\newblock
\APACjournalVolNumPages{A\&A}{608}{}{A143}
\newblock
\PrintBackRefs{\CurrentBib}

\bibitem [\protect \citeauthoryear {%
{Haberl}%
\ \BBA {} {White}%
}{%
{Haberl}%
\ \BBA {} {White}%
}{%
{\protect \APACyear {1990}}%
}]{%
haberl90a}
\APACinsertmetastar {%
haberl90a}%
\begin{APACrefauthors}%
{Haberl}, F.%
\BCBT {}\ \BBA {} {White}, N\BPBI E.%
\end{APACrefauthors}%
\unskip\
\newblock
\APACrefYearMonthDay{1990}{{\APACmonth{09}}}{},
\newblock
\unskip
\newblock
\APACjournalVolNumPages{ApJ}{361}{}{225-234}
\newblock
\PrintBackRefs{\CurrentBib}

\bibitem [\protect \citeauthoryear {%
{Haberl}%
, {White}%
\BCBL {}\ \BBA {} {Kallman}%
}{%
{Haberl}%
\ \protect \BOthers {.}}{%
{\protect \APACyear {1989}}%
}]{%
haberl89a}
\APACinsertmetastar {%
haberl89a}%
\begin{APACrefauthors}%
{Haberl}, F.%
, {White}, N\BPBI E.%
\BCBL {}\ \BBA {} {Kallman}, T\BPBI R.%
\end{APACrefauthors}%
\unskip\
\newblock
\APACrefYearMonthDay{1989}{{\APACmonth{08}}}{},
\newblock
\unskip
\newblock
\APACjournalVolNumPages{ApJ}{343}{}{409-425}
\newblock
\PrintBackRefs{\CurrentBib}

\bibitem [\protect \citeauthoryear {%
{Hickox}%
\ \BBA {} {Vrtilek}%
}{%
{Hickox}%
\ \BBA {} {Vrtilek}%
}{%
{\protect \APACyear {2005}}%
}]{%
hickox05a}
\APACinsertmetastar {%
hickox05a}%
\begin{APACrefauthors}%
{Hickox}, R\BPBI C.%
\BCBT {}\ \BBA {} {Vrtilek}, S\BPBI D.%
\end{APACrefauthors}%
\unskip\
\newblock
\APACrefYearMonthDay{2005}{{\APACmonth{11}}}{},
\newblock
\unskip
\newblock
\APACjournalVolNumPages{ApJ}{633}{}{1064-1075}
\newblock
\PrintBackRefs{\CurrentBib}

\bibitem [\protect \citeauthoryear {%
{Iping}%
\ \BBA {} {Petterson}%
}{%
{Iping}%
\ \BBA {} {Petterson}%
}{%
{\protect \APACyear {1990}}%
}]{%
iping90a}
\APACinsertmetastar {%
iping90a}%
\begin{APACrefauthors}%
{Iping}, R\BPBI C.%
\BCBT {}\ \BBA {} {Petterson}, J\BPBI A.%
\end{APACrefauthors}%
\unskip\
\newblock
\APACrefYearMonthDay{1990}{{\APACmonth{11}}}{},
\newblock
\unskip
\newblock
\APACjournalVolNumPages{A\&A}{239}{}{221-226}
\PrintBackRefs{\CurrentBib}

\bibitem [\protect \citeauthoryear {%
{Klochkov}%
, {Staubert}%
, {Santangelo}%
, {Rothschild}%
\BCBL {}\ \BBA {} {Ferrigno}%
}{%
{Klochkov}%
\ \protect \BOthers {.}}{%
{\protect \APACyear {2011}}%
}]{%
klochkov11a}
\APACinsertmetastar {%
klochkov11a}%
\begin{APACrefauthors}%
{Klochkov}, D.%
, {Staubert}, R.%
, {Santangelo}, A.%
, {Rothschild}, R\BPBI E.%
\BCBL {}\ \BBA {} {Ferrigno}, C.%
\end{APACrefauthors}%
\unskip\
\newblock
\APACrefYearMonthDay{2011}{{\APACmonth{08}}}{},
\newblock
\unskip
\newblock
\APACjournalVolNumPages{A\&A}{532}{}{A126}
\newblock
\PrintBackRefs{\CurrentBib}

\bibitem [\protect \citeauthoryear {%
{Knigge}%
, {Coe}%
\BCBL {}\ \BBA {} {Podsiadlowski}%
}{%
{Knigge}%
\ \protect \BOthers {.}}{%
{\protect \APACyear {2011}}%
}]{%
knigge11a}
\APACinsertmetastar {%
knigge11a}%
\begin{APACrefauthors}%
{Knigge}, C.%
, {Coe}, M\BPBI J.%
\BCBL {}\ \BBA {} {Podsiadlowski}, P.%
\end{APACrefauthors}%
\unskip\
\newblock
\APACrefYearMonthDay{2011}{{\APACmonth{11}}}{},
\newblock
\unskip
\newblock
\APACjournalVolNumPages{Nat}{479}{}{372-375}
\newblock
\PrintBackRefs{\CurrentBib}

\bibitem [\protect \citeauthoryear {%
{Koenigsberger}%
\ \protect \BOthers {.}}{%
{Koenigsberger}%
\ \protect \BOthers {.}}{%
{\protect \APACyear {2006}}%
}]{%
koengisberger06a}
\APACinsertmetastar {%
koengisberger06a}%
\begin{APACrefauthors}%
{Koenigsberger}, G.%
, {Georgiev}, L.%
, {Moreno}, E.%
, {Richer}, M\BPBI G.%
, {Toledano}, O.%
, {Canalizo}, G.%
\BCBL {}\ \BBA {} {Arrieta}, A.%
\end{APACrefauthors}%
\unskip\
\newblock
\APACrefYearMonthDay{2006}{{\APACmonth{11}}}{},
\newblock
\unskip
\newblock
\APACjournalVolNumPages{A\&A}{458}{}{513-522}
\newblock
\PrintBackRefs{\CurrentBib}

\bibitem [\protect \citeauthoryear {%
{Kostka}%
\ \BBA {} {Leahy}%
}{%
{Kostka}%
\ \BBA {} {Leahy}%
}{%
{\protect \APACyear {2010}}%
}]{%
kostka10a}
\APACinsertmetastar {%
kostka10a}%
\begin{APACrefauthors}%
{Kostka}, M.%
\BCBT {}\ \BBA {} {Leahy}, D\BPBI A.%
\end{APACrefauthors}%
\unskip\
\newblock
\APACrefYearMonthDay{2010}{{\APACmonth{09}}}{},
\newblock
\unskip
\newblock
\APACjournalVolNumPages{MNRAS}{407}{}{1182-1187}
\newblock
\PrintBackRefs{\CurrentBib}

\bibitem [\protect \citeauthoryear {%
{Kreykenbohm}%
\ \protect \BOthers {.}}{%
{Kreykenbohm}%
\ \protect \BOthers {.}}{%
{\protect \APACyear {2008}}%
}]{%
kreykenbohm08a}
\APACinsertmetastar {%
kreykenbohm08a}%
\begin{APACrefauthors}%
{Kreykenbohm}, I.%
, {Wilms}, J.%
, {Kretschmar}, P.%
\ et al.\end{APACrefauthors}%
\unskip\
\newblock
\APACrefYearMonthDay{2008}{{\APACmonth{04}}}{},
\newblock
\unskip
\newblock
\APACjournalVolNumPages{A\&A}{492}{}{511-525}
\PrintBackRefs{\CurrentBib}

\bibitem [\protect \citeauthoryear {%
{K{\"u}hnel}%
\ \protect \BOthers {.}}{%
{K{\"u}hnel}%
\ \protect \BOthers {.}}{%
{\protect \APACyear {2017}}%
}]{%
kuehnel17a}
\APACinsertmetastar {%
kuehnel17a}%
\begin{APACrefauthors}%
{K{\"u}hnel}, M.%
, {Rothschild}, R\BPBI E.%
, {Okazaki}, A\BPBI T.%
\ et al.\end{APACrefauthors}%
\unskip\
\newblock
\APACrefYearMonthDay{2017}{{\APACmonth{10}}}{},
\newblock
\unskip
\newblock
\APACjournalVolNumPages{MNRAS}{471}{}{1553-1564}
\newblock
\PrintBackRefs{\CurrentBib}

\bibitem [\protect \citeauthoryear {%
{Leahy}%
\ \BBA {} {Kostka}%
}{%
{Leahy}%
\ \BBA {} {Kostka}%
}{%
{\protect \APACyear {2008}}%
}]{%
leahy08a}
\APACinsertmetastar {%
leahy08a}%
\begin{APACrefauthors}%
{Leahy}, D\BPBI A.%
\BCBT {}\ \BBA {} {Kostka}, M.%
\end{APACrefauthors}%
\unskip\
\newblock
\APACrefYearMonthDay{2008}{{\APACmonth{02}}}{},
\newblock
\unskip
\newblock
\APACjournalVolNumPages{MNRAS}{384}{}{747-754}
\newblock
\PrintBackRefs{\CurrentBib}

\bibitem [\protect \citeauthoryear {%
{Malacaria}%
\ \protect \BOthers {.}}{%
{Malacaria}%
\ \protect \BOthers {.}}{%
{\protect \APACyear {2016}}%
}]{%
malacaria16a}
\APACinsertmetastar {%
malacaria16a}%
\begin{APACrefauthors}%
{Malacaria}, C.%
, {Mihara}, T.%
, {Santangelo}, A.%
, {Makishima}, K.%
, {Matsuoka}, M.%
, {Morii}, M.%
\BCBL {}\ \BBA {} {Sugizaki}, M.%
\end{APACrefauthors}%
\unskip\
\newblock
\APACrefYearMonthDay{2016}{{\APACmonth{04}}}{},
\newblock
\unskip
\newblock
\APACjournalVolNumPages{A\&A}{588}{}{A100}
\newblock
\PrintBackRefs{\CurrentBib}

\bibitem [\protect \citeauthoryear {%
{Manousakis}%
\ \BBA {} {Walter}%
}{%
{Manousakis}%
\ \BBA {} {Walter}%
}{%
{\protect \APACyear {2015}}%
}]{%
manousakis15a}
\APACinsertmetastar {%
manousakis15a}%
\begin{APACrefauthors}%
{Manousakis}, A.%
\BCBT {}\ \BBA {} {Walter}, R.%
\end{APACrefauthors}%
\unskip\
\newblock
\APACrefYearMonthDay{2015}{{\APACmonth{03}}}{},
\newblock
\unskip
\newblock
\APACjournalVolNumPages{A\&A}{575}{}{A58}
\newblock
\PrintBackRefs{\CurrentBib}

\bibitem [\protect \citeauthoryear {%
{Manousakis}%
, {Walter}%
\BCBL {}\ \BBA {} {Blondin}%
}{%
{Manousakis}%
\ \protect \BOthers {.}}{%
{\protect \APACyear {2012}}%
}]{%
manousakis12a}
\APACinsertmetastar {%
manousakis12a}%
\begin{APACrefauthors}%
{Manousakis}, A.%
, {Walter}, R.%
\BCBL {}\ \BBA {} {Blondin}, J\BPBI M.%
\end{APACrefauthors}%
\unskip\
\newblock
\APACrefYearMonthDay{2012}{{\APACmonth{11}}}{},
\newblock
\unskip
\newblock
\APACjournalVolNumPages{A\&A}{547}{}{A20}
\newblock
\PrintBackRefs{\CurrentBib}

\bibitem [\protect \citeauthoryear {%
{Marcu-Cheatham}%
\ \protect \BOthers {.}}{%
{Marcu-Cheatham}%
\ \protect \BOthers {.}}{%
{\protect \APACyear {2015}}%
}]{%
marcu-cheatham15a}
\APACinsertmetastar {%
marcu-cheatham15a}%
\begin{APACrefauthors}%
{Marcu-Cheatham}, D\BPBI M.%
, {Pottschmidt}, K.%
, {K{\"u}hnel}, M.%
\ et al.\end{APACrefauthors}%
\unskip\
\newblock
\APACrefYearMonthDay{2015}{{\APACmonth{12}}}{},
\newblock
\unskip
\newblock
\APACjournalVolNumPages{ApJ}{815}{}{44}
\PrintBackRefs{\CurrentBib}

\bibitem [\protect \citeauthoryear {%
{Mart{\'{\i}}nez-N{\'u}{\~n}ez}%
\ \protect \BOthers {.}}{%
{Mart{\'{\i}}nez-N{\'u}{\~n}ez}%
\ \protect \BOthers {.}}{%
{\protect \APACyear {2017}}%
}]{%
martinez17a}
\APACinsertmetastar {%
martinez17a}%
\begin{APACrefauthors}%
{Mart{\'{\i}}nez-N{\'u}{\~n}ez}, S.%
, {Kretschmar}, P.%
, {Bozzo}, E.%
\ et al.\end{APACrefauthors}%
\unskip\
\newblock
\APACrefYearMonthDay{2017}{{\APACmonth{10}}}{},
\newblock
\unskip
\newblock
\APACjournalVolNumPages{SSRv}{212}{}{59-150}
\newblock
\PrintBackRefs{\CurrentBib}

\bibitem [\protect \citeauthoryear {%
{M{\'e}sz{\'a}ros}%
\ \BBA {} {Nagel}%
}{%
{M{\'e}sz{\'a}ros}%
\ \BBA {} {Nagel}%
}{%
{\protect \APACyear {1985}}%
}]{%
meszaros85a}
\APACinsertmetastar {%
meszaros85a}%
\begin{APACrefauthors}%
{M{\'e}sz{\'a}ros}, P.%
\BCBT {}\ \BBA {} {Nagel}, W.%
\end{APACrefauthors}%
\unskip\
\newblock
\APACrefYearMonthDay{1985}{{\APACmonth{11}}}{},
\newblock
\unskip
\newblock
\APACjournalVolNumPages{ApJ}{298}{}{147-160}
\newblock
\PrintBackRefs{\CurrentBib}

\bibitem [\protect \citeauthoryear {%
{Middleton}%
\ \protect \BOthers {.}}{%
{Middleton}%
\ \protect \BOthers {.}}{%
{\protect \APACyear {2018}}%
}]{%
middleton18a}
\APACinsertmetastar {%
middleton18a}%
\begin{APACrefauthors}%
{Middleton}, M\BPBI J.%
, {Fragile}, P\BPBI C.%
, {Bachetti}, M.%
\ et al.\end{APACrefauthors}%
\unskip\
\newblock
\APACrefYearMonthDay{2018}{{\APACmonth{03}}}{},
\newblock
\unskip
\newblock
\APACjournalVolNumPages{MNRAS}{475}{}{154-166}
\newblock
\PrintBackRefs{\CurrentBib}

\bibitem [\protect \citeauthoryear {%
{Motch}%
, {Pakull}%
, {Soria}%
, {Gris{\'e}}%
\BCBL {}\ \BBA {} {Pietrzy{\'n}ski}%
}{%
{Motch}%
\ \protect \BOthers {.}}{%
{\protect \APACyear {2014}}%
}]{%
motch14a}
\APACinsertmetastar {%
motch14a}%
\begin{APACrefauthors}%
{Motch}, C.%
, {Pakull}, M\BPBI W.%
, {Soria}, R.%
, {Gris{\'e}}, F.%
\BCBL {}\ \BBA {} {Pietrzy{\'n}ski}, G.%
\end{APACrefauthors}%
\unskip\
\newblock
\APACrefYearMonthDay{2014}{{\APACmonth{10}}}{},
\newblock
\unskip
\newblock
\APACjournalVolNumPages{Nat}{514}{}{198-201}
\newblock
\PrintBackRefs{\CurrentBib}

\bibitem [\protect \citeauthoryear {%
{Mushtukov}%
, {Suleimanov}%
, {Tsygankov}%
\BCBL {}\ \BBA {} {Poutanen}%
}{%
{Mushtukov}%
\ \protect \BOthers {.}}{%
{\protect \APACyear {2015}}%
}]{%
mushtukov15a}
\APACinsertmetastar {%
mushtukov15a}%
\begin{APACrefauthors}%
{Mushtukov}, A\BPBI A.%
, {Suleimanov}, V\BPBI F.%
, {Tsygankov}, S\BPBI S.%
\BCBL {}\ \BBA {} {Poutanen}, J.%
\end{APACrefauthors}%
\unskip\
\newblock
\APACrefYearMonthDay{2015}{{\APACmonth{12}}}{},
\newblock
\unskip
\newblock
\APACjournalVolNumPages{MNRAS}{454}{}{2539-2548}
\newblock
\PrintBackRefs{\CurrentBib}

\bibitem [\protect \citeauthoryear {%
{Oskinova}%
, {Feldmeier}%
\BCBL {}\ \BBA {} {Kretschmar}%
}{%
{Oskinova}%
\ \protect \BOthers {.}}{%
{\protect \APACyear {2012}}%
}]{%
oskinova12a}
\APACinsertmetastar {%
oskinova12a}%
\begin{APACrefauthors}%
{Oskinova}, L\BPBI M.%
, {Feldmeier}, A.%
\BCBL {}\ \BBA {} {Kretschmar}, P.%
\end{APACrefauthors}%
\unskip\
\newblock
\APACrefYearMonthDay{2012}{{\APACmonth{04}}}{},
\newblock
\unskip
\newblock
\APACjournalVolNumPages{MNRAS}{421}{}{2820-2831}
\newblock
\PrintBackRefs{\CurrentBib}

\bibitem [\protect \citeauthoryear {%
{Paizis}%
\ \BBA {} {Sidoli}%
}{%
{Paizis}%
\ \BBA {} {Sidoli}%
}{%
{\protect \APACyear {2014}}%
}]{%
paizis14a}
\APACinsertmetastar {%
paizis14a}%
\begin{APACrefauthors}%
{Paizis}, A.%
\BCBT {}\ \BBA {} {Sidoli}, L.%
\end{APACrefauthors}%
\unskip\
\newblock
\APACrefYearMonthDay{2014}{{\APACmonth{04}}}{},
\newblock
\unskip
\newblock
\APACjournalVolNumPages{MNRAS}{439}{}{3439-3452}
\newblock
\PrintBackRefs{\CurrentBib}

\bibitem [\protect \citeauthoryear {%
{Postnov}%
\ \protect \BOthers {.}}{%
{Postnov}%
\ \protect \BOthers {.}}{%
{\protect \APACyear {2015}}%
}]{%
postnov15a}
\APACinsertmetastar {%
postnov15a}%
\begin{APACrefauthors}%
{Postnov}, K\BPBI A.%
, {Gornostaev}, M\BPBI I.%
, {Klochkov}, D.%
, {Laplace}, E.%
, {Lukin}, V\BPBI V.%
\BCBL {}\ \BBA {} {Shakura}, N\BPBI I.%
\end{APACrefauthors}%
\unskip\
\newblock
\APACrefYearMonthDay{2015}{{\APACmonth{09}}}{},
\newblock
\unskip
\newblock
\APACjournalVolNumPages{MNRAS}{452}{}{1601-1611}
\newblock
\PrintBackRefs{\CurrentBib}

\bibitem [\protect \citeauthoryear {%
{Priedhorsky}%
\ \BBA {} {Terrell}%
}{%
{Priedhorsky}%
\ \BBA {} {Terrell}%
}{%
{\protect \APACyear {1983}}%
}]{%
priedhorsky83a}
\APACinsertmetastar {%
priedhorsky83a}%
\begin{APACrefauthors}%
{Priedhorsky}, W\BPBI C.%
\BCBT {}\ \BBA {} {Terrell}, J.%
\end{APACrefauthors}%
\unskip\
\newblock
\APACrefYearMonthDay{1983}{{\APACmonth{10}}}{},
\newblock
\unskip
\newblock
\APACjournalVolNumPages{\apj}{273}{}{709-715}
\newblock
\PrintBackRefs{\CurrentBib}

\bibitem [\protect \citeauthoryear {%
Sato%
\ \protect \BOthers {.}}{%
Sato%
\ \protect \BOthers {.}}{%
{\protect \APACyear {1986}}%
}]{%
sato86a}
\APACinsertmetastar {%
sato86a}%
\begin{APACrefauthors}%
Sato, N.%
, Hayakawa, S.%
, Nagase, F.%
\ et al.\end{APACrefauthors}%
\unskip\
\newblock
\APACrefYearMonthDay{1986}{}{},
\newblock
\unskip
\newblock
\APACjournalVolNumPages{PASJ}{38}{}{731--750}
\PrintBackRefs{\CurrentBib}

\bibitem [\protect \citeauthoryear {%
{Schwarm}%
, {Ballhausen}%
\BCBL {}\ \protect \BOthers {.}}{%
{Schwarm}%
, {Ballhausen}%
\BCBL {}\ \protect \BOthers {.}}{%
{\protect \APACyear {2017}}%
}]{%
schwarm17b}
\APACinsertmetastar {%
schwarm17b}%
\begin{APACrefauthors}%
{Schwarm}, F\BHBI W.%
, {Ballhausen}, R.%
, {Falkner}, S.%
\ et al.\end{APACrefauthors}%
\unskip\
\newblock
\APACrefYearMonthDay{2017}{{\APACmonth{05}}}{},
\newblock
\unskip
\newblock
\APACjournalVolNumPages{A\&A}{601}{}{A99}
\newblock
\PrintBackRefs{\CurrentBib}

\bibitem [\protect \citeauthoryear {%
{Schwarm}%
, {Sch{\"o}nherr}%
\BCBL {}\ \protect \BOthers {.}}{%
{Schwarm}%
, {Sch{\"o}nherr}%
\BCBL {}\ \protect \BOthers {.}}{%
{\protect \APACyear {2017}}%
}]{%
schwarm17a}
\APACinsertmetastar {%
schwarm17a}%
\begin{APACrefauthors}%
{Schwarm}, F\BHBI W.%
, {Sch{\"o}nherr}, G.%
, {Falkner}, S.%
\ et al.\end{APACrefauthors}%
\unskip\
\newblock
\APACrefYearMonthDay{2017}{{\APACmonth{01}}}{},
\newblock
\unskip
\newblock
\APACjournalVolNumPages{A\&A}{597}{}{A3}
\newblock
\PrintBackRefs{\CurrentBib}

\bibitem [\protect \citeauthoryear {%
{Sotani}%
\ \BBA {} {Miyamoto}%
}{%
{Sotani}%
\ \BBA {} {Miyamoto}%
}{%
{\protect \APACyear {2018}}%
}]{%
sotani18a}
\APACinsertmetastar {%
sotani18a}%
\begin{APACrefauthors}%
{Sotani}, H.%
\BCBT {}\ \BBA {} {Miyamoto}, U.%
\end{APACrefauthors}%
\unskip\
\newblock
\APACrefYearMonthDay{2018}{{\APACmonth{08}}}{},
\newblock
\unskip
\newblock
\APACjournalVolNumPages{Phys. Rev. D}{98}{4}{044017}
\newblock
\PrintBackRefs{\CurrentBib}

\bibitem [\protect \citeauthoryear {%
{Staubert}%
\ \protect \BOthers {.}}{%
{Staubert}%
\ \protect \BOthers {.}}{%
{\protect \APACyear {2006}}%
}]{%
staubert06a}
\APACinsertmetastar {%
staubert06a}%
\begin{APACrefauthors}%
{Staubert}, R.%
, {Schandl}, S.%
, {Klochkov}, D.%
, {Wilms}, J.%
, {Postnov}, K.%
\BCBL {}\ \BBA {} {Shakura}, N.%
\end{APACrefauthors}%
\unskip\
\newblock
\APACrefYearMonthDay{2006}{{\APACmonth{06}}}{},
\newblock
{\BBOQ}\APACrefatitle {{Long-term developments in Her X-1: Correlation between
  the histories of the 35 day turn-on cycle and the 1.24 sec pulse period}}
  {{Long-term developments in Her X-1: Correlation between the histories of the
  35 day turn-on cycle and the 1.24 sec pulse period}}.{\BBCQ}
\newblock
\BIn{} F.~{D'Amico}, J.~{Braga}\BCBL {}\ \BBA {} R\BPBI E.~{Rothschild}\
  (\BEDS), \APACrefbtitle {The Transient Milky Way: A Perspective for MIRAX}
  {The Transient Milky Way: A Perspective for MIRAX}\ \BVOL~840, \BPG~65-70.
\newblock
\PrintBackRefs{\CurrentBib}

\bibitem [\protect \citeauthoryear {%
{Tananbaum}%
\ \protect \BOthers {.}}{%
{Tananbaum}%
\ \protect \BOthers {.}}{%
{\protect \APACyear {1972}}%
}]{%
tananbaum72a}
\APACinsertmetastar {%
tananbaum72a}%
\begin{APACrefauthors}%
{Tananbaum}, H.%
, {Gursky}, H.%
, {Kellogg}, E\BPBI M.%
, {Levinson}, R.%
, {Schreier}, E.%
\BCBL {}\ \BBA {} {Giacconi}, R.%
\end{APACrefauthors}%
\unskip\
\newblock
\APACrefYearMonthDay{1972}{{\APACmonth{06}}}{},
\newblock
\unskip
\newblock
\APACjournalVolNumPages{ApJ}{174}{}{L143}
\newblock
\PrintBackRefs{\CurrentBib}

\bibitem [\protect \citeauthoryear {%
{Walter}%
, {Lutovinov}%
, {Bozzo}%
\BCBL {}\ \BBA {} {Tsygankov}%
}{%
{Walter}%
\ \protect \BOthers {.}}{%
{\protect \APACyear {2015}}%
}]{%
walter15a}
\APACinsertmetastar {%
walter15a}%
\begin{APACrefauthors}%
{Walter}, R.%
, {Lutovinov}, A\BPBI A.%
, {Bozzo}, E.%
\BCBL {}\ \BBA {} {Tsygankov}, S\BPBI S.%
\end{APACrefauthors}%
\unskip\
\newblock
\APACrefYearMonthDay{2015}{{\APACmonth{08}}}{},
\newblock
\unskip
\newblock
\APACjournalVolNumPages{Astron. Astrophys. Rev.}{23}{}{2}
\PrintBackRefs{\CurrentBib}

\bibitem [\protect \citeauthoryear {%
Walter%
\ \BBA {} {Zurita-Heras}%
}{%
Walter%
\ \BBA {} {Zurita-Heras}%
}{%
{\protect \APACyear {2007}}%
}]{%
walter07a}
\APACinsertmetastar {%
walter07a}%
\begin{APACrefauthors}%
Walter, R.%
\BCBT {}\ \BBA {} {Zurita-Heras}, J.%
\end{APACrefauthors}%
\unskip\
\newblock
\APACrefYearMonthDay{2007}{{\APACmonth{12}}}{},
\newblock
\unskip
\newblock
\APACjournalVolNumPages{A\&A}{476}{}{335--340}
\PrintBackRefs{\CurrentBib}

\bibitem [\protect \citeauthoryear {%
{Walton}%
\ \protect \BOthers {.}}{%
{Walton}%
\ \protect \BOthers {.}}{%
{\protect \APACyear {2017}}%
}]{%
walton17a}
\APACinsertmetastar {%
walton17a}%
\begin{APACrefauthors}%
{Walton}, D\BPBI J.%
, {Mooley}, K.%
, {King}, A\BPBI L.%
\ et al.\end{APACrefauthors}%
\unskip\
\newblock
\APACrefYearMonthDay{2017}{{\APACmonth{04}}}{},
\newblock
\unskip
\newblock
\APACjournalVolNumPages{ApJ}{839}{}{110}
\newblock
\PrintBackRefs{\CurrentBib}

\bibitem [\protect \citeauthoryear {%
White%
, Swank%
\BCBL {}\ \BBA {} Holt%
}{%
White%
\ \protect \BOthers {.}}{%
{\protect \APACyear {1983}}%
}]{%
white83a}
\APACinsertmetastar {%
white83a}%
\begin{APACrefauthors}%
White, N\BPBI E.%
, Swank, J\BPBI H.%
\BCBL {}\ \BBA {} Holt, S\BPBI S.%
\end{APACrefauthors}%
\unskip\
\newblock
\APACrefYearMonthDay{1983}{{\APACmonth{07}}}{},
\newblock
\unskip
\newblock
\APACjournalVolNumPages{ApJ}{270}{}{711--734}
\PrintBackRefs{\CurrentBib}

\bibitem [\protect \citeauthoryear {%
{Wolff}%
\ \protect \BOthers {.}}{%
{Wolff}%
\ \protect \BOthers {.}}{%
{\protect \APACyear {2016}}%
}]{%
wolff16a}
\APACinsertmetastar {%
wolff16a}%
\begin{APACrefauthors}%
{Wolff}, M\BPBI T.%
, {Becker}, P\BPBI A.%
, {Gottlieb}, A\BPBI M.%
\ et al.\end{APACrefauthors}%
\unskip\
\newblock
\APACrefYearMonthDay{2016}{{\APACmonth{11}}}{},
\newblock
\unskip
\newblock
\APACjournalVolNumPages{ApJ}{831}{}{194}
\newblock
\PrintBackRefs{\CurrentBib}

\end{thebibliography}

\end{document}